\documentclass[oldversion,letter]{aa}  
\usepackage{epsfig}
\usepackage{natbib}
\usepackage{txfonts}
\usepackage{graphicx}

\def\m2s2{\hbox{\,m$^{2}$\,s$^{-2}$}} 
\def\Msun{\hbox{$M_{\odot}$}}             

\def\Me{\hbox{$M_{\oplus}$}}
\def\Re{\hbox{$R_{\oplus}$}}
\def\degr{\hbox{$^\circ$}}

\def\chisq{\mbox{$\chi^2$}}

\begin{document}

\title{Limits to the planet candidate GJ 436c\thanks{Based on observations taken with the Telescopio Carlos S\'anchez (TCS) of the Observatorio del Teide, operated by the Instituto de Astrof\'\i sica de Canarias.} }

\author{Alonso, R. \inst{1}
\and Barbieri, M. \inst{1}
\and Rabus, M. \inst{2}
\and Deeg, H.J. \inst{2}
\and Belmonte, J.A. \inst{2}
\and Almenara, J.M. \inst{2}
}

\offprints{\email{roi.alonso@oamp.fr}}

\institute{Laboratoire d'Astrophysique de Marseille, UMR 6110, CNRS/Universit\'e de Provence, Traverse du Siphon, 13376 Marseille, France
\and
Instituto de Astrof\'\i sica de Canarias, E-38205 La Laguna, Spain
}

\date{Received / Accepted }

\abstract
{We report on H-band, ground-based observations of a transit of the hot Neptune GJ~436b. Once combined to achieve sampling equivalent to archived observations taken with Spitzer, our measurements reach comparable precision levels. We analyze both sets of observations in a consistent way, and measure the rate of orbital inclination change to be of 0.02$\pm$0.04\degr in the time span between the two observations (253.8~d, corresponding to 0.03$\pm$0.05\degr yr$^{-1}$ if extrapolated). This rate allows us to put limits on the relative inclination between the two planets by performing simulations of planetary systems, including a second planet, GJ~436c, whose presence has been recently suggested (Ribas et al. 2008). The allowed inclinations for a 5\Me\, super-Earth GJ~436c in a 5.2~d orbit are within $\sim$7\degr of the orbit of GJ~436b; for larger differences the observed inclination change can be reproduced only during short sections ($<$50\%) of the orbital evolution of the system. The measured times of three transit centers of the system do not show any departure from linear ephemeris, a result that is only reproduced in $<$1\% of the simulated orbits. Put together, these results argue against the proposed planet candidate GJ~436c.}

\titlerunning{Limits to the planet candidate GJ 436c}

\authorrunning{Alonso et al.}

\maketitle

\section{Introduction}
\label{sec:intro}
Among the known transiting exoplanets, GJ~436b is by far the smallest and least massive exoplanet, being to date the only transiting hot Neptune. With a mass of 23~\Me, it orbits a M star with 0.44~\Msun, in a slightly eccentric orbit (e=0.15), with a period of 2.64~d. The exoplanet GJ~436b was discovered by a radial velocity survey \citep{butler04}, and in the same work the authors reported a non-detection of photometric transits. Three years later, \cite{gillon_disc} announced the discovery of transits of this object, which provided the first (and currently the only) measurement of the radius and density of a Neptune-sized transiting exoplanet. Its radius is currently reported to be of around 4.22~\Re \, \citep{torres07}, and according to models of icy and rocky planets (\citealt{seager,fortney}), it is thought that the planet has an H-He envelope (\citealt{gillon_disc,deming_sp}). Several authors have signaled that the significant non-zero eccentricity is an effect of a further companion in the system that excites the eccentricity, otherwise the system should be circularized in relatively short timescales (\citealt{maness07,deming_sp}).

The low inclination of the orbit (86.3\degr), the non-detection of transits by \cite{butler04}, and an analysis of a low-significance peak in the residuals of the radial velocity data led \cite{ribas08} to suggest the presence of a 5~\Me \,super-Earth (GJ~436c) orbiting with a period of 5.2~d. An important argument of this hypothesis is the change in the inclination of GJ~436b induced by the gravitational interactions with GJ~436c. Under the assumption that there was a $\sim$0.3 deg variation in the orbital inclination between 2004 and 2007 --which would have made the transit non-detection by \cite{butler04} and the detection by \cite{gillon_disc} compatible-- \cite{ribas08} suggested that the transits observed in 2008 should last $\sim$2 minutes longer, due to a rate of inclination change of roughly 0.1\degr yr$^{-1}$.

In this work, we report ground-based H-band observations of a transit of GJ~436b observed in March 2008, and perform a reanalysis of the Spitzer 8~$\mu$m data reduced by \cite{gillon_sp}. We use different techniques aimed at measuring the rate of inclination change on GJ~436b, and provide constraints on the inclination and masses of the proposed planet GJ~436c.
\section{Observations and data reduction}
\label{sec:obs}
Data were collected at the 1.52~m Telescopio Carlos S\'anchez (TCS) telescope, operated by the Instituto de Astrof\'\i sica de Canarias at the Observatorio del Teide, during the night of March 8, 2008. We used a H filter and the detector CAIN-II\footnote{http://www.iac.es/telescopes/cain/cain.html}, a NICMOS 3 technology chip, which employs a 256$\times$256 array of HgCdTe elements sensible to the range 1-2.5 microns. The selected \emph{W} widefield optical configuration of the detector provides an image scale of 1$\arcsec$/pixel. 

Due to the high brightness of the star, we severely defocussed the image in order not to saturate the star with a 1.5~s exposure time. This resulted in ring-shaped images of the star with an outer diameter of $\sim$26$\arcsec$ and an inner diameter of $\sim$7$\arcsec$. The flux of the star was thus spread into $\sim$500 pixels, which helps to minimize the effects of bad pixels, tracking, and atmospheric seeing changes. We selected a part of the detector with few hot or dead pixels and, to achieve a good stability of the system, we employed no dithering pattern. To account for hot pixels inside the target's PSF, we moved the telescope every 1-2 hours to a close zone of the sky with no stars and took 50 exposures. Pixels with a low or zero sensitivity (dead pixels) were located in images of the dome under a slight artificial illumination. They represent $\sim$3\% of the detector. A Fowler readout mode was employed, performing 8 readouts per image and rejecting 2 of them. Taking into account the observing overheads of typically 1~s, we obtained a total of $\sim$8000 exposures in a single night.

\begin{figure}
\begin{center}
\epsfig{file=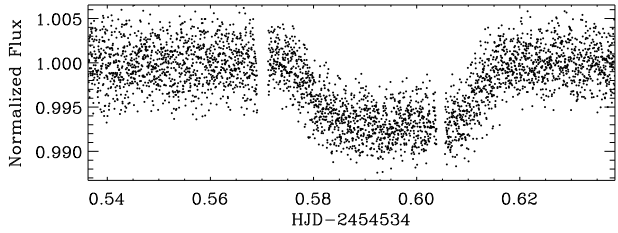,width=9cm}
\caption{Normalized light curve of GJ~436 on the night of March 8, 2008. The dispersion of the data out of transit phases is 0.0023, with a median time sampling of 2.35~s. } 
\label{fig:fig1}
\end{center}
\end{figure}

To extract the flux of the star, we first interpolated the dead pixels we found as explained above. We used the average images of each set of 50 background exposures to subtract the hot pixel contribution. The flux was summed inside a circular aperture of 20$\arcsec$ in radius and centered on the star, and the background level was estimated from an annulus of 22$\arcsec$ and 34$\arcsec$ with the same center. The integrated flux of the star in a single 1.5~s exposure was $\sim$2.6$\times$10$^7$ e$^-$. We restricted our analysis to the phases close to the primary transit of GJ~436b. The star exhibited a long term variation of only $\sim$1\% during the night, which was well corrected by a parabolic fit to the parts outside the transit. In two moments of the night (around HJD-2454534 = 0.5736 and 0.603) we observed a fast decrease and increase of the flux, in a 20-30~s timescale, probably caused by slight vignetting of the dome. We discarded these points for the rest of the analysis. The final normalized light curve that was used for the analysis described in the next section, containing 3600 points, is plotted in Fig.~\ref{fig:fig1}. The remarkably low noise level attained can be attributed to both the smooth variations of the atmospheric transparency in the H filter, as well as to the decision not to use a dithering pattern, which could have emphasized the detector inhomogeneities.

\section{Analysis}
\label{sec:ana}
In order to evaluate the rate of orbital inclination change for GJ~436b, we modeled the TCS and the 8$\mu$m Spitzer data in a consistent way. Both observations of transits are separated by 253.8~d. For the transit modeling, we used the formalism of \cite{gim06}, which is also valid for eccentric orbits, such as the case of GJ~436b. The best fit was found by a minimization of the \chisq\, using the commonly used AMOEBA algorithm (\citealt{press}). To estimate the error of each of the fitted parameters, we performed 100 fits in different sets of light curves, which were constructed by subtracting the best-fit model, shifting circularly the vectors of the residuals and their errors by a random number, and again adding the best-fit model. With this bootstrapping procedure, we allow for the possible low-frequency structure of the residuals due to uncorrected systematic effects.
\subsection{TCS data}
We grouped the TCS data in bins with a width of $\sim$0.00012 in phase, corresponding to $\sim$27~s, which is approximately the same time sampling as the Spitzer data set. The errors were estimated as the standard dispersion of the points inside each bin, divided by the square root of the number of points inside each bin. The solution that provides the best fit is presented in Table~\ref{tab:parpl}. The fitted limb darkening coefficients are in good agreement with the values reported by \cite{claret} for a star with T$_{eff}$=3500~K, log$g$=4.5.  
\begin{figure}
\begin{center}
\epsfig{file=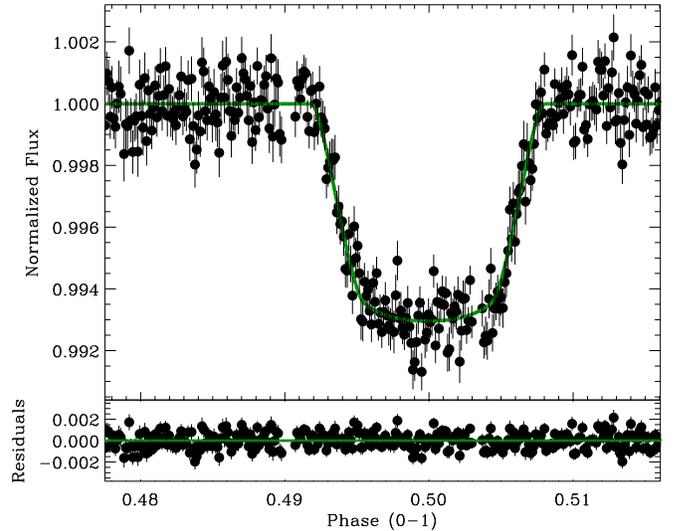,width=9cm,angle=0}
\caption{TCS H-band phased light curve of GJ~436 and best-fit model (green line), and the residuals from the best-fit model (bottom). The bin size corresponds to $\sim$27~s, and the 1-sigma error bars have been estimated from the dispersion of the points inside each bin. The standard deviation of the residuals is 0.00077.}
\label{fig:fig2}
\end{center}
\end{figure}
\subsection{Spitzer data}
\begin{figure}
\begin{center}
\epsfig{file=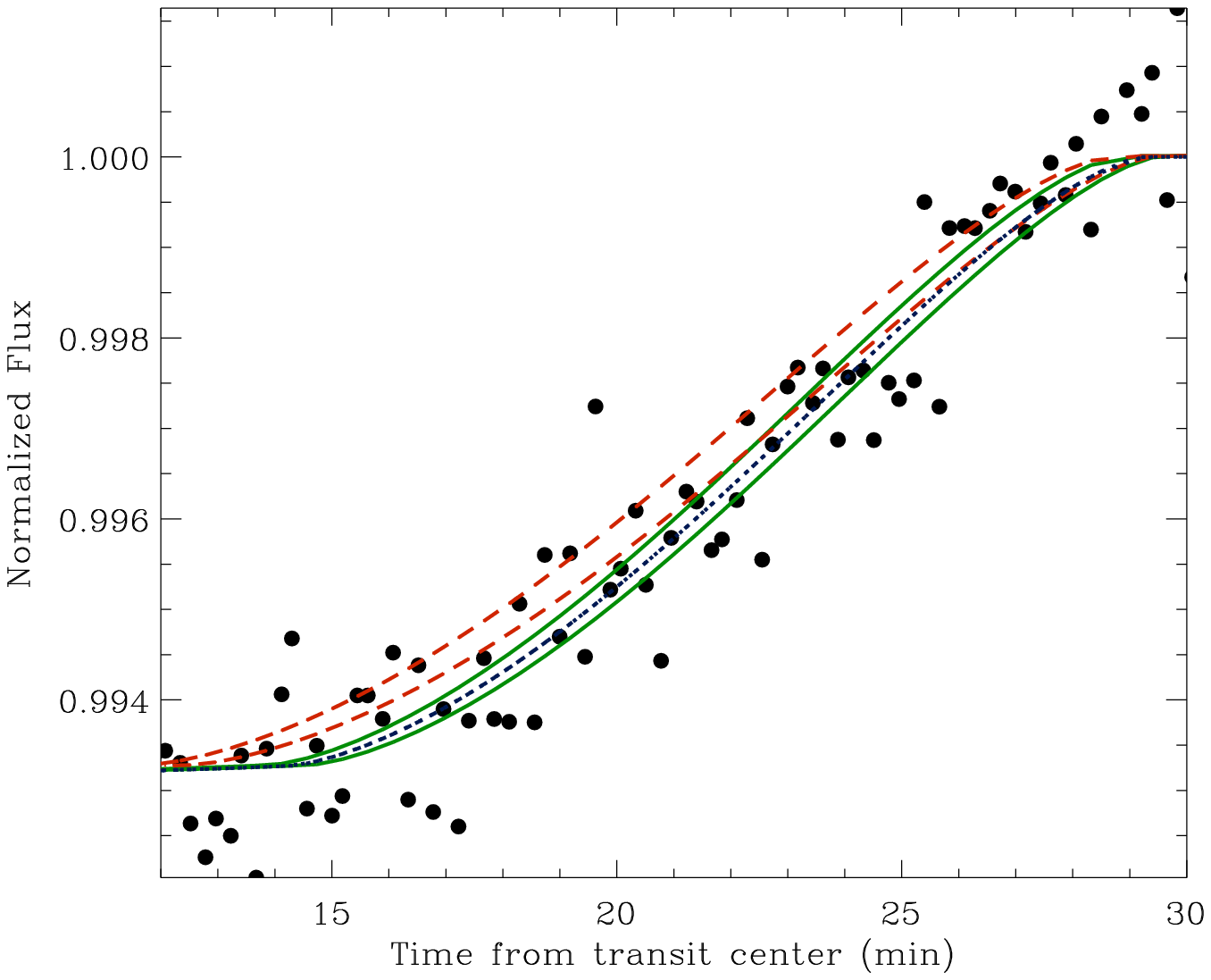,width=9cm,angle=0}
\caption{The region of ingress/egress of the Spitzer data of \cite{gillon_sp}, and three different models: 1) Our best-fit model, with an inclination of 86.54\degr and taking into account the eccentric orbit (solid green line). 2) Our best-fit model, for a circular orbit (dotted blue line). The inclination is 86.3\degr, in perfect agreement with \cite{torres_param}. 3) A model with an inclination of 86.3\degr and taking into account the eccentric orbit (dashed red line). Note that for eccentric orbits, the times from the transit center to the transit ingress and egress are slightly different, as a consequence of the change in orbital velocity during the transit, making the lines appear doubled in this plot.}
\label{fig:fig3}
\end{center}
\end{figure}

One transit of GJ~436b has been observed with Spitzer at 8~$\mu$m (\citealt{deming_sp,gillon_sp}). We analyzed the set of data reduced by \cite{gillon_sp}, obtaining the results in Table~\ref{tab:parpl}. We note that we obtain a different inclination than that reported by \cite{torres_param}, which is lower by 0.2\degr. By modeling the transit with a zero eccentricity, we obtain an inclination in perfect agreement with the inclination reported by \cite{torres_param}, and $\sim$0.4\degr higher than the solution proposed by \cite{gillon_sp}. In Fig.~\ref{fig:fig3}, we plot the three different fits, which shows the importance of properly including the eccentricity of the orbit in the transit modeling. Due to the low-frequency noise in the parts of total eclipse, we obtained unrealistic values of limb-darkening if we also fitted for these parameters. We thus fixed these parameters to those reported by \cite{gillon_sp}.
\subsection{Comparison of both data sets}
One of the predicted effects of a planet with similar characteristics as that proposed by \cite{ribas08} is the variation of the orbital inclination (and thus, the total duration of the transits) of GJ~436b. Measuring the \emph{absolute} orbital inclination using transits is a challenging task, due to the effects of uncertainties on the limb darkening coefficients and the possible effects of active regions on the star. A simpler approach consists in measuring \emph{relative} orbital inclination changes of the same object by the more precise --and less affected by the uncertainties on the limb darkening coefficients-- determination of the total transit duration.
As a first, simple method, we evaluated the total duration of the two transits and their transit centers with a fit to a trapezoidal function, where the fitted parameters were the transit center, duration, depth, and ingress time. The fit was performed with a Levenberg-Marquardt algorithm (\citealt{press}). The duration of the transit observed with Spitzer is estimated, with this approach, to be 58.7$\pm$0.7~m, while the TCS best-fitted trapezoid lasts 59.2$\pm$0.8~m. The centers of the transits were used to determine the ephemeris of Table~\ref{tab:parpl}. The error bars in the timing of the transit centers were estimated from the dispersion of the bootstrap fits described in the previous section, and thus include the influence of uncorrected systematics in both sets of data. Alternatively, we can also compute the total transit duration from the value of the fitted $\theta _1$; in this case, it is necessary to perform the conversion between the circular orbital phase used intrinsically in the \cite{gim06} formulation and the observed orbital phase. Using this method, we obtain a duration of 60.5$\pm$1.0~m and 61.0$\pm$0.7~m for Spitzer and TCS transits, respectively. We attribute the slight difference between the absolute values using both methods to the effect of the limb darkening, considered to be zero when performing trapezoidal fits. The relative difference in total duration $\Delta$D of the transits using both methods is in perfect agreement. This difference in duration can be converted to a difference in orbital inclination $\Delta i_{obs}$ of 0.02$\pm$0.04\degr between the two observations, or 0.03$\pm$0.05\degr yr$^{-1}$ if we assume the $\Delta i_{obs}$ behaves linearly in these timescales. 

\begin{table}
\begin{minipage}[t]{\columnwidth}
\begin{center}
\caption[]{Parameters of the GJ~436 fits, and associated one-sigma errors.}  
\renewcommand{\footnoterule}{}  
\begin{tabular}{lrr}
\hline\hline
& Value& Error\\
$P$ [d]&  2.64390 &0.00003\\
$T_{c}$ Discovery [HJD]\footnote{From \cite{gillon_disc}} &2454222.616&0.001\\
$T_{c}$ Spitzer [HJD] &  2454280.78153& 0.00028\\
$T_{c}$ TCS [HJD] &  2454534.59584& 0.00015\\
$e$ & 0.15 &(fixed) \\
$w$ & 343\degr& (fixed) \\
\hline
\multicolumn{3}{c}{\emph{TCS H-band photometry}} \\
$\theta_{1}$\footnote{Phase of transit ingress in the reference system defined by \cite{gimepelayo}} & 0.00801&0.00008\\
$k=R_p/R_s$ &  0.0841&0.0011\\
$i$ [deg]& 86.78&0.21\\
$u_+$\footnote{$u_+=u_a+u_b$, with $u_a$ and $u_b$ the coefficients of a quadratic law for the limb darkening.} & 0.43&0.17\\
$u_-$\footnote{$u_-=u_a-u_b$} & -0.69&0.13\\
\hline
\multicolumn{3}{c}{\emph{Spitzer photometry}} \\
$\theta_{1}$ & 0.00791&0.00012\\
$k=R_p/R_s$ &  0.0835&0.0014\\
$i$ [deg]& 86.54&0.13\\
$u_+$ & 0.14&(fixed)\\
$u_-$ & -0.05&(fixed)\\
\hline
$\Delta D$ [min] & 0.5 & 0.8 \\
$\Delta i_{obs}$ [deg] & 0.02 & 0.04 \\
$\Delta i$ [deg yr$^{-1}$] & 0.03 & 0.05 \\

\hline
\label{tab:parpl}
\end{tabular}
\end{center}
\end{minipage}
\end{table}
\section{Discussion}
\begin{figure}
\begin{center}
\epsfig{file=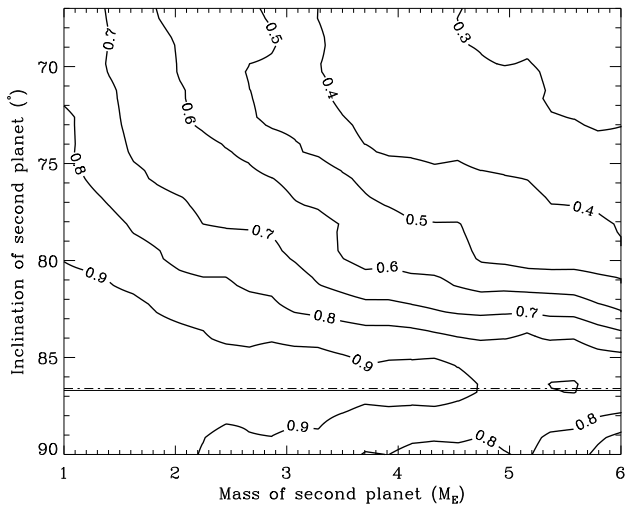,width=9cm,angle=0}
\caption{The fraction of time that the simulated planetary systems verified the $\Delta i_{obs}$ restrictions summarized in Table~\ref{tab:parpl}, as a function of the mass of the second planet and its orbital inclination. The horizontal lines mark the observed inclination of GJ~436b (solid line) and the lower limit to the inclination of GJ~436c to produce transits (dashed line).}
\label{fig:fig4}
\end{center}
\end{figure}

In order to put physical limitations on a second planet from the observational constraint of the previous section, we assembled numerical models of two-body planetary systems. We defined the initial osculating Keplerian orbital elements\footnote{The observed Keplerian orbital elements at a particular epoch. See, for instance, \cite{roy}} of the GJ~436b from the values of \cite{torres07} and chose the orbital plane of this planet as the reference plane of the system. The elements of GJ~436c were defined according to the values proposed by \cite{ribas08}. The parameters for the planet $c$ were chosen randomly in the intervals resumed in Table~\ref{t:sim}; the angular parameters not present in this table were chosen randomly for the two planets. The evolution of the system was calculated using the RADAU integrator \citep{1985dcto.proc..185E}, which ensures a proper reproduction of the close encounters between the planets. We performed a total of 5\,000 simulations, following the evolution of the system for 1400~yr, which are $\sim$2$\times$10$^5$ orbits of GJ~436b, and $\sim$10$^5$ orbit of the outer planet. The orbital parameters of GJ~436b were recorded every 253.81~days, which is the difference in time between TCS and Spitzer observations, to be able to perform comparisons between the $\Delta i$ obtained from the simulations and the observed $\Delta i_{obs}$.
\begin{table}
\centering
\caption{Mass of the star \object{GJ 436}, initial osculating Keplerian elements for GJ~436b and ranges for GJ~436c.}
\begin{tabular}{lr|lr}
\hline\hline
GJ~436      &                      &           &                       \\
M$_\star$ &    0.452~\Msun&           &                       \\
\hline
GJ~436b  &                      & GJ~436c &                       \\
$m_b$     &    23.17~\Me & $m_c$     &    1--6~\Me\\
$a_b$     &    0.02872 AU        & $a_c$     &0.044--0.046 AU        \\
$e_b$     &      0.15            & $e_c$     &    0--0.2             \\
$i_b$     &       0\degr     & $i_c$     &    0--20\degr      \\
$\omega_b$&      343\degr    & $\omega_c$&  222--308\degr    \\
\end{tabular}
\label{t:sim}
\end{table}
For each of the simulations containing N time stamps, we calculated the fraction $f=N(\Delta i<\Delta i_{obs})/N$ of time stamps in which the change of the inclination of GJ~436b was within 1~$\sigma$ of the measured $\Delta i_{obs}$. We explored the variation of $f$ as a function of the mass and inclination of the second planet. We created a regular grid in this parameter space, and computed the mean value of $f$ in each cell. The result is represented in Fig.~\ref{fig:fig4}. As expected from orbital interactions, massive planets induce relatively large inclination changes unless their inclination is close to the inclination of GJ~436b. Thus, the change in inclination sampled by only two points will be less than $\Delta i_{obs}$ only in a very small fraction $f$. A 5~\Me\, planet has a value of $f$ bigger than 0.5 only when its inclination is within $\sim$7\degr of that of GJ~436b, and planets as less massive as 3~\Me\, produce comparable values of $f$ if their inclination difference is around 15\degr. 

Furthermore, the suggested period for GJ~436c is close to a 2:1 mean-motion resonance, what would induce transit time variations (TTV) of the order of several minutes (\citealt{holman,agol}). For two particular cases (5\Me\, and 2\Me), and assuming a difference in orbital inclinations of 5\degr, we calculated the times of minimum projected separation between the centers of the star and the planet GJ~436b, to track the evolution of the TTV for 10~yr. In order to compare with the observed values of center times (Table~\ref{tab:parpl}), we randomly chose a start epoch $E_0$, and calculated the standard deviation of the TTVs at $E_0$,$E_{-21}$, and $E_{+96}$ (there are 21 orbits between the discovery transit of Gillon et al. 2007b and the transit observed with Spitzer, and 96 between this last transit and the transit observed with TCS). We repeated the process 10\,000 times, and calculated the fraction in which this value was smaller than the standard deviation of the observed TTV (12~s).
For a 5\Me\, planet, this fraction was of 0.5\%, rising to 5\% for the 2\Me\, case. Hence, there is a very low probability that the non detection of TTV in three epochs is in agreement with the existence of a 5\Me\, companion in a period close to the 2:1 resonance.
%
%

However, the low-measured value of $\Delta i_{obs}$, compatible with zero within 1-$\sigma$, and the absence of significant TTV both strongly argue against the proposed 4.8\Me\, planet in a 5.2~d orbit. Further measurements of transit epochs could serve to restrict the few remaining systems that produce apparent TTVs of the order of the observed values.


Additionally, we have shown that ground-based observations with a 1.5~m class telescope in the near infrared can attain precisions that are of the order of those obtained with Spitzer at higher wavelengths. These kind of observations are specially suited for transit modeling and TTV studies.
\begin{acknowledgements}
These observations have been funded by the Optical Infrared Coordination Network (OPTICON), a major international collaboration supported by the Research Infrastructures Programme of the European CommissionÕs Sixth Framework Programme. MR would like to thank the European Association for Research in Astronomy (EARA) for their support by the EARA - Marie Curie Early Stage Training fellowship. We are grateful to the Observatorio del Teide staff for precious help during the observations.
\end{acknowledgements}
%

\end{document}